# Electronic structure and chemical bonding anisotropy investigation of wurtzite AlN

M. Magnuson[1], M. Mattesini[2], C. Höglund[1], J. Birch[1] and L. Hultman[1]

[1] *Department of Physics, IFM, Thin Film Physics Division, Linköping University, SE-58183 Linköping, Sweden.*

[2] *Departamento de Física de la Tierra, Astronomía y Astrofísica I, Universidad Complutense de Madrid, E-28040, Spain.*

**Abstract**

The electronic structure and the anisotropy of the Al - N π and σ chemical bonding of wurtzite AlN has been investigated by bulk-sensitive total fluorescence yield absorption and soft x-ray emission spectroscopies. The measured N $K$, Al $L_1$, and Al $L_{2,3}$ x-ray emission and N 1$s$ x-ray absorption spectra are compared with calculated spectra using first principles density-functional theory including dipole transition matrix elements. The main N 2$p$ - Al 3$p$ hybridization regions are identified at -1.0 to -1.8 eV and -5.0 to -5.5 eV below the top of the valence band. In addition, N 2$s$ - Al 3$p$ and N 2$s$ - Al 3$s$ hybridization regions are found at the bottom of the valence band around -13.5 eV and -15 eV, respectively. A strongly modified spectral shape of Al 3$s$ states in the Al $L_{2,3}$ emission from AlN in comparison to Al metal is found, which is also reflected in the N 2$p$ - Al 3$p$ hybridization observed in the Al $L_1$ emission. The differences between the electronic structure and chemical bonding of AlN and Al metal are discussed in relation to the position of the hybridization regions and the valence band edge influencing the magnitude of the large band gap.

## 1 Introduction

Wurtzite aluminum nitride (*w*-AlN) is an interesting group-III type of semiconductor material with many important technological applications. It has a large direct band gap (6.2 eV), exhibit piezoelectricity, high thermal conductivity, mechanical strength and low electron affinity [1, 2]. This makes it an attractive material for optoelectronic device applications, e.g., LEDs in the visible (violet-blue) and UV-range, field emitters for flat panel displays, high-speed transistors, and microelectronic devices at high temperatures and high powers. Providing greater performance characteristics than classical III-V semiconductors in many applications, the wide band gap AlN, GaN and SiC materials, are considered the third generation of semiconductor materials in the microelectronics industry. GaN and AlN form ternary $Ga_xAl_{1-x}N$ alloys, with $0 \leq x \leq 1$. By choosing a suitable composition of the alloy, a tuning to smaller bandgaps than for AlN corresponding to longer emission wavelengths of the optoelectronic devices is enabled. However, for AlN there is a lack of detailed experimental data concerning the basic electronic structure and hybridization regions of the relatively strong Al-N chemical bonds, in particular, at the bottom of the valence band. In the wurtzite crystal structure, there are two types of Al-N bonds ($\pi$ and $\sigma$) with slightly different bond lengths.

Previous experimental studies of *w*-AlN include x-ray photoelectron spectroscopy (XPS) [3] for probing the occupied states and soft x-ray absorption spectroscopy (SXA) to measure the anisotropy of the unoccupied N $2p$ states [4]. Nonresonant soft x-ray emission (SXE) has also been utlizied at the *K*-edges of N and Al [4, 5, 6] of AlN. However, for Al *K* emission measurements using a grating spectrometer at hard x-rays, the agreement between the experimental spectra and the calculated density of states was not satisfying. On the other hand, for the shallower Al $L_1$ and $L_{2,3}$ edges of AlN, the hybridization regions, anisotropy and chemical bonding to N, has remained unclear.

In this work, we investigate the bulk electronic structure and the influence of hybridization and chemical bonding between the constituent atoms of single-crystalline *w*-AlN in comparison to single-crystal Al metal, using bulk-sensitive and element-specific SXE. Angle-dependent SXA spectroscopy in bulk-sensitive total fluorescence yield mode (SXA-TFY) is also utilized to probe the unoccupied states at the N $1s$ edge. The occupied valence-band contributions of the N $2p$, Al $3d$, Al $3p$ and Al $3s$ states have been measured by recording the *K*-edge spectra of N and the $L_1$ and $L_{2,3}$ emission of Al. The SXE and SXA-TFY techniques - around the N $1s$, Al $2s$, and Al $2p$ absorption thresholds - are much more bulk sensitive than electron-based spectroscopic techniques. Due to the involvement of both valence and core levels, the corresponding difference in energies of the emission lines and their dipole selection rules, each kind of atomic element can be probed separately. This enables to extract both elemental and chemical bonding information of the electronic structure of the valence and conduction bands. The SXE and SXA-TFY spectra are interpreted in terms of calculated spectra using symmetry selective partial density of states (pDOS) weighted by the dipole transition matrix elements. In SXA-TFY at the N $1s$ edge, we observe a strong polarization dependence corresponding to different weights of the $\pi$ and $\sigma$ bonds which agrees well with the predicted anisotropy of the unoccupied conduction-band states. As shown in comparison to related binary nitride systems, the N $2p$ - Al $3p$, N $2s$ - Al $3p$ and N $2s$ - Al $3s$ hybridizations affect the valence-band edge and influence the magnitude of the large band gap.

## 2 Experimental

### 2.1 Deposition and growth of AlN

A ~ 250 nm thick single crystal stoichiometric AlN(0001) thin film was grown epitaxially by ultrahigh vacuum (UHV) reactive magnetron sputtering onto a polished MgO(111) substrate, $10\times10\times0.5$ mm$^3$ in size, at 300 $^o$C. The deposition was carried out in an UHV vacuum chamber with a base pressure of $1.0\times10^{-8}$ Torr from an unbalanced type II magnetron with a 75 mm diameter Al target (99.999 %). Prior to deposition the substrate was cleaned in ultrasonic baths of trichloroethylene, acetone and isopropanol and blown dry in dry $N_2$. This was followed by degassing in the chamber for 1 h at 900 $^o$C before ramping down to the deposition temperature at 300 $^o$C. The substrate potential was set to be -30 V. During deposition the Ar and $N_2$ partial pressures were kept at 1.0 mTorr and 3.0 mTorr, respectively, while the magnetron power was set to 270 W. The crystal structure was characterized with $\Theta/2\Theta$ diffraction (XRD) scans in a Philips Bragg-Brentano diffractometer using Cu-$K_\alpha$ radiation. Figure 1 (top), shows single crystalline $w$-AlN, where the main 0002 and 0004 reflections are observed at 36.12$^o$ and 76.64$^o$ below the MgO substrate peaks. XRD from bulk Al is shown below with the main reflection at 44.71$^o$ corresponding to fcc Al(100). Additional small reflections are found at 38.47$^o$, 65.07$^o$ and 78.13$^o$ with much lower intensity indicating minor contributions from other crystalline orientations.

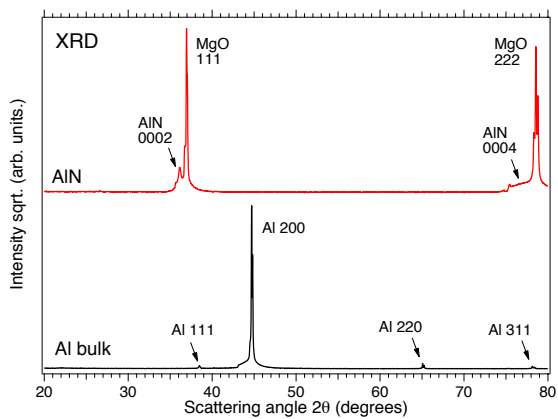

**Figure 1:** (Color online) X-ray diffractograms (XRD) from the AlN thin film sample in comparison to bulk Al(100).

### 2.2 X-ray emission and absorption measurements

The SXE and SXA measurements were performed at the undulator beamline I511-3 at MAX II (MAX-lab National Laboratory, Lund University, Sweden), comprising a 49-pole undulator and a modified SX-700 plane grating monochromator [7]. The SXE spectra were measured with a high-resolution Rowland-mount grazing-incidence grating spectrometer [8] with a two-dimensional multichannel detector with a resistive anode readout. The N $K$ SXE spectra were recorded using a spherical grating with 1200 lines/mm of 5 m radius in the first order of diffraction. The Al $L_1$ and $L_{2,3}$ spectra were recorded using a grating with 300 lines/mm, of 3 m radius in the first order of diffraction. The SXA spectra at the N 1*s* edges were measured at both normal and 20*o* grazing incidence with 0.08 eV resolution in total fluorescence yield mode (SXA-TFY). During the N *K*, Al $L_1$ and $L_{2,3}$ SXE measurements, the resolutions of the beamline monochromator were 0.45, 0.2 and 0.1 eV, respectively. The N *K*, Al $L_1$ and Al $L_{2,3}$ SXE spectra were recorded with spectrometer resolutions of 0.42, 0.3 and 0.06 eV, respectively. All measurements were performed with a base pressure lower than $5 \times 10^{-9}$ Torr. In order to minimize self-absorption effects [9, 10], the angle of incidence was 20*o* from the surface plane during the SXE measurements. The x-ray photons were detected parallel to the polarization vector of the incoming beam in order to minimize elastic scattering.

# 3 *Ab initio* calculation of soft x-ray absorption and emission spectra

The SXA and SXE spectra were calculated with the WIEN2K code [11] employing the density-functional [12, 13] augmented plane wave plus local orbital (APW+lo) computational scheme [14]. The APW+lo method expands the Kohn-Sham (KS) orbitals in atomic-like orbitals inside the muffin-tin atomic spheres and plane waves in the interstitial region. The KS-equations were solved for the wurtzite (B4) AlN phase (*w*-AlN) using the Wu-Cohen generalized gradient approximation (WC-GGA) [15, 16] for the exchange-correlation potential. Our electronic band structure calculations were carried out by employing the optimized WC-GGA a-lattice parameter and *c*/*a* ratio. For the hexagonal close-packed lattice of *w*-AlN, we obtained a=3.117 Å and *c*/*a*=1.615, which are in excellent agreement with the reported experimental values (*a*=3.111 Å and *c*/*a*=1.601 [17]) and other theoretical calculations [18, 19, 20, 21]. The fitting of the energy *versus* unit cell volume data set was performed by using the Birch-Murnaghan [22] equation of state and gave a bulk modulus of 207.7 GPa (B=4.158), which is also in a remarkable agreement with the experimental value of 207.9 GPa [17].

The computations of SXE spectra were carried out within the so-called *final-state rule* [23], where no core-hole was created at the photo-excited atom. Theoretical emission spectra were computed at the converged ground-state density by multiplying the orbital projected partial density of states (pDOS) with the energy dependent dipole matrix-element between the core and the valence band states [24]. The comparison with the experimental spectra was achieved including an instrumental broadening in the form of Gaussian functions. The final state lifetime broadening was also accounted for by a convolution with an energy-dependent Lorentzian function with a broadening increasing linearly with the distance from the top of the valence band.

The SXA spectra were calculated by including core-hole effects, thus specifically considering a crystal potential created from a static screening of the core-hole by the valence electrons. Such a self-consistent-field potential was generated in a 2×2×2 hexagonal supercell of 32 atoms (16 independent) containing one core-hole on the probing atomic specie. The electron-neutrality of the system was kept constant by means of a negative background charge. By this procedure we explicitly include the excitonic coupling between the screened core-hole and the conduction electrons, which is an important requisite for simulating absorption edges in large band-gap semiconductors [25]. The Al-2$s$, -2$p$ and N-1$s$ absorption edges were then generated by weighting the empty pDOS with the dipole matrix-element between the core and the conduction band states. The instrumental and lifetime broadening was applied according to the same methodology used for the SXE spectra. Finally, in order to study possible anisotropy effects in the $w$-AlN phase, the theoretical SXE and SXA spectra were also computed for the N $p_z$ and N $p_{xy}$ components separately.

# 4 Results

## 4.1 N 1$s$ x-ray absorption and *K* emission

Figure 2 (top-right curves) shows SXA-TFY spectra following the N 1$s \rightarrow$ 2$p$ dipole transitions. By measuring in both normal and grazing incidence geometries, the unoccupied *2$p_{xy}$* ($\sigma^*$) and *2$p_z$* ($\pi^*$) orbitals are probed, respectively. The $w$-AlN crystalline film has the $c$-axis oriented parallel to the surface normal. Consequently, for SXA in grazing incidence geometry when the *E*-vector is almost parallel to the $c$-axis, the unoccupied out-of-plane (2$p_z$) single $\pi^*$ orbitals are preferably excited and projected at the N 1$s$-edge. On the other hand, for SXA at normal incidence geometry, when the *E*-vector is orthogonal to the $c$-axis, the four tetrahedrally coordinated unoccupied $\sigma^*$ orbitals in the *2$p_{xy}$* plane are preferably excited and probed. Note that the N 1$s$ SXA spectra which were measured both at normal and grazing incidence exhibit a strong intensity variation with three distinguished peaks each corresponding to the unoccupied $\pi^*$ and $\sigma^*$ orbitals with different weights. At normal incidence, three peak structures are observed with *2$p_{xy}$* $\sigma^*$ symmetry at 400.7 eV, 403.5 eV and 406.5 eV. At grazing incidence, the intensity of the central peak has emerged into a shoulder of the third peak. The measured anisotropy of the unoccupied *2$p_{xy}$* and *2$p_z$* orbitals is in excellent agreement with the calculated SXA spectra shown at the bottom of Fig. 2. The SXA spectra were also used to determine the photon energy of the first absorption peak maximum at the N 1$s$ threshold used for the SXE measurements (vertical tick).

The N *K* SXE spectra following the N $2p \rightarrow 1s$ dipole transitions of *w*-AlN, excited at 400.9 and 430.0 eV photon energies are shown in the left part of Fig. 2. In near normal exit geometry, the intensity of the SXE spectra of *w*-AlN are dominated by the occupied N $2p_{xy}$ (σ) orbitals by the N $2p \rightarrow 1s$ dipole transitions. The main $2p_{xy}$ σ peak is very sharp and located at -1.0 eV on the relative energy scale at the bottom relative to the top of the valence band. A well-separated and prominent low-energy shoulder is located at -5.5 eV below the top of the valence band. Note that the Rayleigh *elastic* scattering peak of *w*-AlN at the resonant excitation energy (400.9 eV) is very weak. The magnitude of the elastic scattering primarily depends on the experimental geometry and the surface roughness. The apparent elastic peak (re-emission) is known to be strong in correlated materials with localized 3*d* and 4*f* electrons as observed for rare earth materials [26] where multiplet calculations are usable. Calculated N *K* SXE spectra using DFT with projected $2p_{xy}$ and $2p_z$ symmetries are shown at the bottom of Fig. 2. The calculated SXE spectra are in satisfactory agreement with the experimental data although the -5.5 eV low-energy shoulder is located at -4.9 eV in the calculations. The intensity of the calculated $2p_z$ ($\pi$) SXE spectrum is lower than for the $2p_{xy}$ ($\sigma$) spectrum with an additional substructure between the main peak and the shoulder. We also note that in *w*-AlN, the $\pi$ and $\sigma$ bond distances between Al and N are comparable, where the single $\pi$ bond along the c-axis is slightly shorter (1.907 Å) in comparison to the three N-Al bonds with $\sigma$ symmetry (1.910 Å).

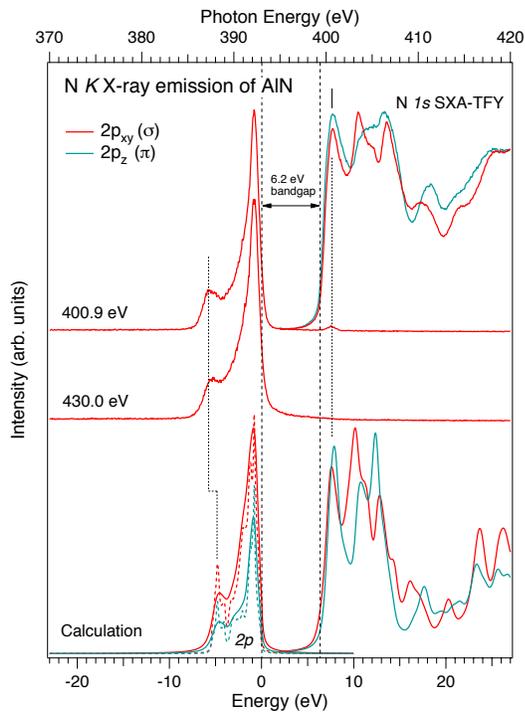

**Figure 2:** (Color online) top right: N 1*s* SXA-TFY spectra and left: resonant and nonresonant N *K* SXE spectra of *w*-AlN. The SXA-TFY spectra were normalized by the step edge below and far above the N 1*s* absorption thresholds. All spectra were plotted on a photon energy scale (top) and a relative energy scale (bottom) with respect to the top of the valence band. The SXE spectra were resonantly excited at the first N 1*s* absorption peak maximum at 400.9 eV and nonresonant far above the N 1*s* absorption threshold at 430.0 eV. The excitation energy (400.9 eV) of the resonant SXE spectrum is indicated by the vertical tick in the SXA spectrum and by the small elastic recombination peak at 400.9 eV. To account for the experimental bandgap of 6.2 eV, the calculated SXA spectra were rigidly shifted by +2.24 eV. The calculated N 2*p* charge occupation of *w*-AlN is 2.286e.

The estimated band gap between the valence and conduction band edges of SXE and SXA is 6.2 ± 0.2 eV, in good agreement with optical absorption measurements [27]. Our WC-GGA-DFT calculations of the computed direct band gap [$E_g(\Gamma-\Gamma)$] amounts to 3.96 eV, which is 2.24 eV smaller than the experimental value (6.2 eV). This general deficiency of both LDA- and GGA-DFT methods has been extensively studied in the past [28, 29], and it is now well-accepted that it can be partly overcome by introducing a specific on-site Coulomb interaction [30]. However, for semiconductors, the main effect comes from the electron exchange-correlation discontinuity through the band gap [31]. Due to the underestimation of the *ab initio* band gap, the calculated SXA spectra were rigidly shifted by +2.24 eV in Fig. 2.

## 4.2 Al $L_1$ and $L_{2,3}$ x-ray emission

Figure 3 shows Al $L_1$ (top panel) and Al $L_{2,3}$ (bottom panel) SXE spectra of AlN and single-crystal Al metal [32], following the $3p \rightarrow 2s$ and $3s, 3d \rightarrow 2p_{3/2,1/2}$ dipole transitions, respectively. The measurements were made nonresonantly at 140 eV and 110 eV photon energies. Calculated spectra with the dipole projected pDOS and appropriate core-hole lifetime broadening are shown below the measured spectra. Note the ~ 3 eV chemical shift of the top of the valence band in AlN in comparison to Al metal. The general agreement between experiment and theory is better for the $L_1$ emission involving spherically symmetric 2s core levels than for the $L_{2,3}$ emission involving 2p core levels. The $L_1$ fluorescence yield is much lower than the $L_{2,3}$ yield making the $L_1$ measurements more demanding in terms of extensive beamtime. The main Al $L_1$ emission peak in *w*-AlN is found at -1.5 eV on the common energy scale is due to Al 3p orbitals hybridizing with the N 2p orbitals. On the contrary, the weak $L_1$ emission of pure Al metal is very broad and flat without any narrow peak structures in the whole energy region, in agreement with our calculated $L_1$ spectrum of the $2p_{xy}$ $\sigma$ orbitals shown below. In comparison to the spectrum of Al metal, the $L_1$ spectral structure of *w*-AlN is largely focused to a specific energy region (-1.5 eV), as a consequence of Al 3p hybridization with the bonding N 2p orbitals between -1 to -4 eV in Fig. 2. The shoulder at -5 to -5.5 eV is a signature of additional Al 3p - N 2p hybridization and bonding corresponding to the -5.5 eV shoulder in the N *K* SXE spectrum in Fig. 2. In addition, a peak feature is observed at the bottom of the valence band at -13.5 eV due to N 2s - Al 3p hybridization. The calculated Al $L_1$ spectra are generally in good agreement with the experiment, although the calculated peak positions are at somewhat higher energy and the difference increases with the distance from the top of the valence band.

The bottom panel of Fig. 3 shows measured Al $L_{2,3}$ SXE spectra of w-AlN due to $3s3d \rightarrow 2p_{3/2,1/2}$ dipole transitions. The peak at -5.0 eV is assigned predominantly to Al $3s$ states whereas the peak maximum at -1.8 eV has contributions from both Al $3s$ and $3d$ states. For Al metal, the partially occupied $3d$-contribution slowly increases with energy and becomes maximal near the top of the valence band. The situation is different in solid AlN as the $3d$ states hybridize with the N $2sp$ states throughout the entire valence band which stabilize the system. A charge transfer occurs to the empty $3d$-states as the Al-N bonds form in the solid. The strong Al $3d$ - N $2p$ hybridization may be one of the reasons for the larger band gap of AlN (6.2 eV) in comparison to GaN (3.4 eV) [33]. Generally, the electronic structure of the Al atoms is strongly modified by the neighboring N atoms. In particular, this is evident when comparing to the Al $L_{2,3}$ spectrum of

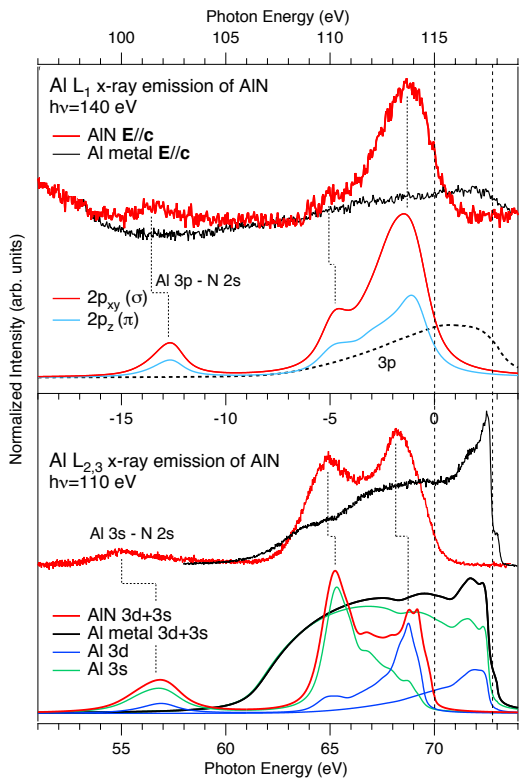

**Figure 3:** (Color online) Experimental and calculated Al $L_1$ and Al $L_{2,3}$ SXE spectra of w-AlN in comparison to Al(100) spectra. The experimental spectra were excited nonresonantly at 140 eV and 110 eV, respectively. The vertical dotted line indicates the valence band edge. A common energy scale with respect to the $E_F$ is indicated in the middle. For comparison of the peak intensities and energy positions, the integrated areas of the experimental and calculated spectra of the two systems were normalized to the calculated Al $3p$ and $3d+3s$ charge occupations of w-AlN: (Al $3s$: 0.524e, $3p$: 0.701e, $3d$: 0.282e) and pure Al metal ($3s$: 0.592e, $3p$: 0.526e, $3d$: 0.090e). The area for the $L_2$ component was scaled down by the experimental branching ratio and added to the $L_3$ component.

Al metal where a very sharp peak has its maximum at -0.22 eV below the $E_F$ in Al metal due to an edge resonance [34]. The small shoulder at +0.24 eV above $E_F$ of Al metal is due to Al $L_2$ emission. The 2$p$ spin-orbit splitting is 0.46 eV, slightly larger than our calculated *ab initio* spin-orbit splitting of 0.44 eV. The Al 2$p$ spin-orbit splitting is not resolved in $w$-AlN. In contrast to the $L_{2,3}$ SXE spectrum of pure Al metal, the Al $L_{2,3}$ spectrum of $w$-AlN has a strongly modified spectral weight towards lower emission energy. As the main peak has its maximum at -1.8 eV below the top of the valence band and a low-energy peak at -5.0 eV, it is an indication of hybridization with the N 2$p$ orbitals in this energy region. For the calculated Al $L_{2,3}$ SXE spectra, the $3s,3d \rightarrow 2p_{3/2,1/2}$ matrix elements play an important role for the spectral functions as the intensity from the pDOS at the top of the valence band is enhanced and at the bottom reduced. However, the main disagreement between the experimental edge resonance of Al metal and the one-electron calculation especially at threshold can be attributed to many-body effects, addressed in the Mahan - Noziéres - De Dominicis MND theory of edge singularities [35, 36].

# 5 Discussion

From Figs. 2 and 3, we distinguished two main hybridization regions giving rise to the N 2$p$ - Al 3$p$ bonding and band formation at -1.0 to -1.8 eV and -5.0 to -5.5 eV below the top of the valence band. In addition, Al 3$p$ - N 2$s$ and Al 3$s$ - N 2$s$ hybridization regions are found at the bottom of the valence band around -13.5 eV and -15 eV, respectively. The Al 3$p$ - N 2$s$ and Al 3$s$ - N 2$s$ hybridization and covalent bonding are both deeper in energy from the top of the valence band than the N 2$p$ - Al 3$p$ hybridization indicating relatively strong bonding although the intensities are lower.

Comparing the N $K$ and Al $L_1$, $L_{2,3}$ SXE spectra of $w$-AlN to the binary ScN, TiN, and the ternary Sc$_3$AlN [37] and Ti$_2$AlN [32] nitride compounds, we firstly note that there is a shift of the bonding N 2$p$ orbitals relative to the upper valence band edge. In $w$-AlN, the N $K$ SXE peak appear at -1.0 eV with a well-separated low-energy shoulder at -5.5 eV. For ScN and Sc$_3$AlN, the corresponding main peak and shoulder appear at lower energy at -4 and -6 eV [37]. For TiN and Ti$_2$AlN, the main peak and the shoulder are located at -4.8 and -6 eV, respectively [32]. This implies a shift of the metal 3$d$ pDOS towards lower energy which also affects the spectral distributions of the Al $L_1$ and $L_{2,3}$ spectra. In all cases, Al 3$p$ and the Al 3$s$ hybridization peaks are located about -1.5 eV and -5 eV below the top of the valence band. However, for $w$-AlN, we also observe an Al 3$p$ low-energy shoulder at -5.0 eV to -5.5 eV and a peak at -13.5 eV hybridizing with the N 2$s$ orbitals at the bottom of the valence band. In addition, the Al 3$d$ peak at -1.8 eV dominates over the 3$s$ peak in $w$-AlN and a peak corresponding to Al 3$s$ - N 2$s$ hybridization is observed at -15 eV at the bottom of the valence band. Secondly, TiN, Ti$_2$AlN and Sc$_3$AlN are all good conductors while ScN has a band gap of approximately 1.6 eV [37] which makes this material interesting as a VUV optical material. Thirdly, comparing the electronic structure of *w*-AlN with fcc Al metal, it is clear that the physical properties and the electronic structure are strongly affected by the hybridization with the N atoms. The change of the *p* - *d* hybridization regions and the band filling of the electronic structure affecting and chemical shifting the valence band edge is the main reason for the large 6.2 eV band gap observed in *w*-AlN. This is also evident when comparing the anion and cation *s*, *p*, *d* charge occupations of *w*-AlN (Al 3*s*: 0.524e, 3*p*: 0.701e, 3*d*: 0.282e, N 2*p*: 2.286e) with those of ScN (Sc 3*s*+4*s*: 2.017e, 3*p*: 5.774e, 3*d*: 0.675e, N 2*p*: 2.173e) [37] and TiN (Ti 3*s*+4*s*: 2.060e, 3*p*: 5.920e, 3*d*: 1.532e, N 2*p*: 2.180e) [32]. The 3*d* orbitals generally stabilize the crystal structure of *w*-AlN in comparison to pure Al metal as less 3*d* states are located at the very top of the valence band edge but at lower energy.

Substitution and alloying of Al in AlN to Ga in GaN would also change the charge in the *sp* valence bands. However, as in the case of Ge [38], the shallow 3*d* core level, at -19 eV below the top of the valence band in Ga will effectively interact and withdraw charge density from the *sp* valence band which is expected to influence the hybridization and chemical bonding in the material. This will be the subject of further investigations.

# 6 Conclusions

In summary, we have investigated the electronic structure and chemical bonding of wurtzite AlN in comparison to Al metal. The combination of bulk sensitive and element selective soft x-ray absorption spectra in fluorescence yield mode, soft x-ray emission spectroscopy and, electronic structure calculations show that in wurtzite AlN, the main N 2*p* - Al 3*p* hybridization and bond regions appear -1.0 eV to -1.8 eV and -5.0 to -5.5 eV below the top of the valence band. In addition, N 2*s* - Al 3*p* and N 2*s* - Al 3*s* hybridization regions are found at the bottom of the valence band at -13.5 eV and -15 eV, respectively. Our measured Al $L_{2,3}$ emission spectra of wurtzite AlN as compared to pure Al metal shows a significant chemical shift of the main 3*d* intensity from the top of the valence band edge to a hybridization region at -1.8 eV. This signifies an effective transfer of charge from the Al 3*d* and 3*s* orbitals to the Al 3*p* and N 2*p* orbitals stabilizing the crystal structure. A strong polarization dependence in the unoccupied N 2*p* states is also identified as due to different weights of the empty $2p_{xy}$ - $\sigma^*$ and $2p_z$ - $\pi^*$ orbitals. The major materials property effect of the change of the electronic structure, charge occupation and the chemical bond regions in comparison to Al metal is the opening of the wide band gap of 6.2 eV in wurtzite AlN.

# 7 Acknowledgements

We would like to thank the staff at MAX-lab for experimental support. This work was supported by the Swedish Research Council Linnaeus Grant LiLi-NFM, the Göran Gustafsson Foundation, the Swedish Strategic Research Foundation (SSF), Strategic Materials Reseach Center on Materials Science for Nanoscale Surface Engineering (MS$^2$E). One of the authors (M. Mattesini) wishes to acknowledge the Spanish Ministry of Science and Technology (MCyT) for financial support through the *Ramón y Cajal* program.